# Electronegativity and chemical hardness of the elements under pressure

Xiao Dong[1*], Artem R. Oganov[2*], Haixu Cui[3], Xiang-Feng Zhou[4], and Hui-Tian Wang [5]

[1]Key Laboratory of Weak-Light Nonlinear Photonics and School of Physics, Nankai University, Tianjin 300071, China

[2] Skolkovo Institute of Science and Technology, Skolkovo Innovation Center, 3 Nobel Street, Moscow 143026, Russia

[3] College of Physics and Materials Science, Tianjin Normal University, Tianjin 300387, China

[4] Center for High Pressure Science, State Key Laboratory of Metastable Materials Science and Technology, School of Science, Yanshan University, Qinhuangdao 066004, China

[5] Collaborative Innovation Center of Advanced Microstructures, Nanjing University, Nanjing, 210093, China

*Corresponding author. Email: xiao.dong@nankai.edu.cn (X. D.), a.oganov@skoltech.ru (A. R. O.)

**Abundant evidence has shown the emergence of exotic chemical phenomena under pressure, including the formation of unexpected compounds and strange crystal structures. In many cases, there is no convincing explanation for these phenomena and there are virtually no chemical rules or models capable of predicting or even rationalizing these phenomena. Here we calculate, as a function of pressure, two central chemical properties of atoms, electronegativity and chemical hardness, which can be seen as the first and second-order chemical potentials. Mulliken electronegativity, which equals minus the chemical potential of the electron relative to the vacuum, is appropriately modified – instead of taking the vacuum (impossible under high pressure), we take the homogeneous electron gas as reference. We find that for most elements, chemical hardness and electronegativity decrease with pressure, consistent with pressure-induced metallization. Furthermore, we discover that pressure-induced *s-d* orbital transfer makes Ni, Pd and Pt "pseudo-noble-gas" atoms with a closed d-shell configuration, and the elements preceding them (Fe and especially Co, Rh, Ir) electron acceptors, while the elements right after them (Cu, Ag, Zn, Cd, for example) become highly electropositive. We show the explicative and predictive power of our electronegativity and chemical hardness scales under pressure.**

## I. Introduction

Recent theoretical and experimental investigations have established that pressure greatly affects chemical properties of the elements[1]. For example, pressure increases the reactivity of noble gases: e.g., according to theory and experiment xenon oxides become thermodynamically stable at moderate pressures (> 75 GPa) [2,3], Na and He react at 113 GPa to form an unusual compound $Na_2He$ [4], and stable stoichiometric compounds of helium, such as $Na_2HeO$ [4]; $CaF_2He$, $MgF_2He$ [5]; $SiO_2He$ [6]; $H_2OHe$ [7,8], have been predicted at moderate pressures. Caesium displays multivalent states (such as $Cs^{III}$ and $Cs^{V}$) in the predicted pressure-stabilized $CsF_n$ ($n$ > 1) compounds [9,10]. Furthermore, under pressure, unexpected sodium chlorides, such as $Na_3Cl$ and $NaCl_3$, become stable [11]. Such compounds with stoichiometries that cannot be anticipated from atomic valences become ubiquitous under pressure and include highest-temperature superconductors known to date, such as $LaH_{10}$ [12-15] ; $H_3S$ [16,17] ; $ThH_{10}$ and $ThH_9$ [18,19]; $YH_6$ [20,21].

To put these cases of dramatic changes of chemistry into a general and predictive system, the simplest approach is to determine how the essential chemical properties of the atoms change under pressure. The most important properties are (1) the electronic configuration, (2) size, and (3) electronegativity and chemical hardness. For (1), it is well known that under pressure the orbitals with higher angular momentum become favorable – hence, atoms typically undergo *s-p* and *s-d* transitions[1,22]. For (2), atomic sizes (volumes) decrease and can be easily tabulated, at any pressure. As to (3), electronegativity and chemical hardness can be expected to be highly non-trivial. Recently, an attempt [23] was made to calculate atomic electronegativities as a function of pressure using the single atom Hamiltonian with eXtreme Pressure Polarizable Continuum Model (XP-PCM), but electronegativities therein did not contain the essential $PV$-term. Here we reassess electronegativities of all elements from H to Cm as a function of pressure, properly taking all pressure-related effects (including the $PV$-term) into account, and also calculating the chemical hardnesses of the elements.

## II. Theoretical analysis

Ionization potential and electron affinity are atomic properties of paramount importance. Ionization potential $E_i$ is defined as the energy of reaction $A \rightarrow A^{+} + e$ and the electron affinity $E_a$ is *minus* the energy of reaction $A + e \rightarrow A^{-}$. The ionization potential is defined as $E_i = E(1) - E(0)$ and the electron affinity is $E_a = E(0) - E(-1)$, where $E(0)$, $E(1)$ and $E(-1)$ indicate the atomic energy with 0, +1 and −1 charges, respectively. Then Mulliken electronegativity $\chi$ and chemical hardness $\eta$ are defined as:

$$\chi = (E_i + E_a)/2 \text{ and } \eta = \left(E_i - E_a\right)/2 \quad (1)$$

Expanding the energy $E$ of a given atom in powers of the number $N$ of electrons as $E(N) = a + bN + cN^2 + o(N^3)$, [24] we find that its first derivative, which is the chemical potential of the electron in the atom, equals to minus Mulliken electronegativity, $dE/dN\big|_{N=0} = \mu = -\chi$. This electronegativity quantifies the ability of an atom to attract and retain electrons. The second derivative is the chemical hardness as $\eta = c = \frac{1}{2} d^2E/dN^2\big|_{N=0}$, which describes the resistance of an atom to a change of its electronic state.

With Koopmans' theorem [25], one finds that the chemical potential is the midpoint between Highest Occupied Molecular Orbital (HOMO) and Lowest Unoccupied Molecular Orbital (LUMO), and the chemical hardness equals to half of the HOMO-LUMO gap. Solid-state analogs of Mulliken electronegativity and of the chemical hardness are the work function and the band gap, respectively, and indeed a reasonable correlation of the electronegativity and work function is well known [26,27].

At zero pressure, an electron added to a neutral atom feels only a modest nuclear attraction, so $E_a$ is much smaller than $E_i$ - therefore, electronegativity and chemical hardness have similar trends: active metals have low electronegativity and low chemical hardness, and active non-metals have high electronegativity and high chemical hardness. Noble gases are anomalous because of their practically zero $E_a$ and very high $E_i$.

To discuss physical reasons behind the change of chemical properties of the atoms with pressure, we have to consider the response of atoms to pressure. Obviously, pressure does not change nuclear charges and the numbers of electrons, but it affects the spatial distribution of the electrons through the change of atomic volume $V$. We should consider not only the shrinking of the wave function and its effects on the kinetic and potential energy, but also switch to the use of the enthalpy $H = E + PV$ (rather than energy $E$) as the relevant thermodynamic potential.

To get the enthalpies at high pressure, we use the "helium matrix method" [28]: a sufficiently large (3×3×3) supercell of the perfect He fcc structure where we replace one atom of He by one atom of the element of interest and give that atom a charge of 0, −1 or +1, and relax the structure at any pressure of interest.

Before discussing the results, we make four comments:

1. Mulliken's definition is inapplicable at high pressure and must be appropriately modified, introducing a proper reservoir of electrons and using the relevant thermodynamic potential (enthalpy instead of energy). The standard definition of the ionization potential $E_i$ is $\mathrm{AHe}_{107} \rightarrow \mathrm{AHe}_{107}^{+} + \mathrm{e}$ and the electron affinity $E_a$ is defined as minus the energy of reaction $\mathrm{AHe}_{107} + \mathrm{e} \rightarrow \mathrm{AHe}_{107}^{-}$ - in both cases "e" denotes an isolated electron in the vacuum. At non-zero pressure there is no vacuum, and the electron will have to be in some reservoir where it will have finite volume and the corresponding $PV$ contribution. Different reservoirs of the electron could be used, but we choose the homogeneous electron gas, the idealized metallic system with exactly known enthalpy as a function of pressure [29,30]. Our new electronegativities have the meaning of minus the chemical potential of the electron in the atom, *relative to the electron gas at the same pressure*. This is a useful feature: for example, an atom with a negative electronegativity would lose its electron in favor of the electron gas. Note that our definition of electronegativity at zero pressure differs from Mulliken's by -2.11 eV, which equals to the energy of the electron gas (per one electron) at zero pressure; the remaining difference from experimental values is about -0.84 eV and is due to the use of approximate DFT functional, and effects of the He-matrix. Here we use the PBE-GGA functional (see Methods), and note that this functional gives exact energy for the electron gas, i.e. is compatible with the reference we use. We also note that the contribution of the reservoir is important only for the electronegativity; it cancels out for the chemical hardness, which is therefore reservoir-independent and equivalent to the traditional definition.

2. It is advisable to use monopole and quadrupole corrections [31] for periodic charged systems to guarantee fast convergence with respect to the supercell size.

3. Whenever not stated explicitly otherwise, the data in this paper are calculated with spin polarization and spin-orbit coupling (SOC). In most cases, the differences are minor (Fig. S1). The largest difference is the increase of the chemical hardness upon inclusion of spin polarization for elements with half-filled shells, such as nitrogen ($p^3$). For a few $d$-block elements, spin polarization introduces some changes. Pressure induces electronic transitions in some atoms, and this also affects electronegativities and chemical hardnesses – however, as these are based on enthalpies (rather than just internal energies as in Rahm et al[23]), they display no discontinuities, but only changes of the slope at the transition. This allows one to confidently interpolate between the pressure points at which we determined these properties. As for SOC, it makes only minor changes for heavy elements, see Fig S1.

4. Compared with experimental data[32], the errors in our results for $E_i$ and $E_a$ at zero pressure are small (Fig. S2a and b), and for most elements they are smaller than 0.5 eV and mostly come from errors of the electronic structure method. An additional source of minor errors is that in some extreme cases, the He matrix cannot localize the charges near the impurity atom: for elements with nearly zero electron affinity, such as alkali metals or noble gases, the extra electron added to the neutral atom to some extent delocalizes over the He matrix. Likewise, in the calculation of the ionization enthalpy of He and F the positive charge is partially delocalized over the He matrix. So, for a few elements, the computed affinity enthalpy is overestimated or their ionization enthalpy is underestimated. However, as these elements still have different abilities to attract the electron, we do obtain different values, correctly representing the trends and qualitatively meaningful. For most elements the results are quantitative.

5. Here we report calculated electronegativities at 0 GPa (actual calculations were done at 1 atm = $10^{-4}$ GPa), 50 GPa, 200 GPa, 500 GPa. For several elements values at other pressures were calculated as well.

## III. Results

### 1. Trends for the electronegativity

It is reassuring that at ambient pressure many metals have electronegativities close to zero – i.e. the chemical potential of the electron in them is about the same as in the electron gas. With increasing pressure, the repulsive effect of atomic cores makes electronegativities of metal atoms strongly negative. The most electropositive elements at zero pressure are alkali metals, followed by heavy alkali earths and some f-elements, in the order of increasing electronegativity: Cs→K→Fr→Rb→Ra→Pu→Ba. With pressure this order changes, at 500 GPa it is: Na,Mg→Ag→Ac→Li→In→Cs→Cu,Rb. Amazingly, coinage metals Cu, Ag, Au have joined the ranks of the most electropositive elements. Equally unexpected is that elements of different groups are now closer to each other in properties than to members of their own groups – for example, at 500 GPa, Na and Mg are much more similar than Na and K.

The most electronegative elements at zero pressure, as expected, are the halogens, noble gases, oxygen and nitrogen, in the order of decreasing values: F→He→Ne→Cl→O→Br→Ar→N→H→I. At 500 GPa fluorine is still the most electronegative element: F→O→Cl→H→N→He→Ne (these are the only elements with positive electronegativity at 500 GPa), followed by heavy noble gases and halogens.

From the thermodynamic relation $d(\Delta H)/dP = \Delta V$, it follows that whether the enthalpy increases or decreases, depends on the sign of the volume difference $\Delta V$. We can naturally define the affinity volume $V_a$ and the ionization volume $V_i$, induced by attracting or losing one electron, and both are always negative. The increase or decrease of electronegativity depends on whether the sum $V_{fe}+(V_i+V_a)/2$ is positive or negative ($V_{fe}$ is the volume of the free electron at given pressure, and is always positive), these values are given in Fig. S5 and Table S3.

**2. Trends for the chemical hardness**

Chemical hardness is a measure of convexity of the $E(N)$ function and quantifies stability of the neutral state of the atom. Our results show that chemical hardness for most elements decreases with pressure, reaching much lower values than at zero pressure and (since the chemical hardness within Koopmans' theorem approximation is equal to the HOMO-LUMO gap) indicating the tendency to pressure-induced metallization. Also, for all elements at least up to 500 GPa chemical hardness is positive, i.e. neutral atom X is stable against charge disproportionation (i.e. reaction $2X = X^{+}+X^{-}$ is unfavorable).

From values in Supplementary Table S2 we see that most elements with chemical hardness below 3.75 eV are metals (except Si, Ge and Te) and most elements with chemical hardness above 3.75 eV are non-metals (except Be, Mg, Cd, Zn, Hg, Mn, Rh, Pd, Hf), and most of the exceptions are within 1 eV of this borderline.

At zero pressure, noble gases have the highest chemical hardness, in the obvious order of decreasing values: Ne→He→Kr→Xe→Rn, followed by hydrogen, fluorine and other elements. While hardnesses of most elements decrease with pressure, Ni-group elements show a peculiar trend of increasing hardness, and at 500 GPa the highest chemical hardness is possessed by both noble gases and Ni-group elements, in the order of decreasing values: Ne→He→Ar→Kr→Pd→Xe→Pt→Ni, then followed by hydrogen and other elements. High chemical hardness means particular stability of the electronic structure of these elements. The fact that Ne, rather than He, has the highest chemical hardness, is perfectly consistent with the prediction that Ne has the highest pressure of metallization (208.4 TPa)[33], even higher than He (32.9 TPa)[34].

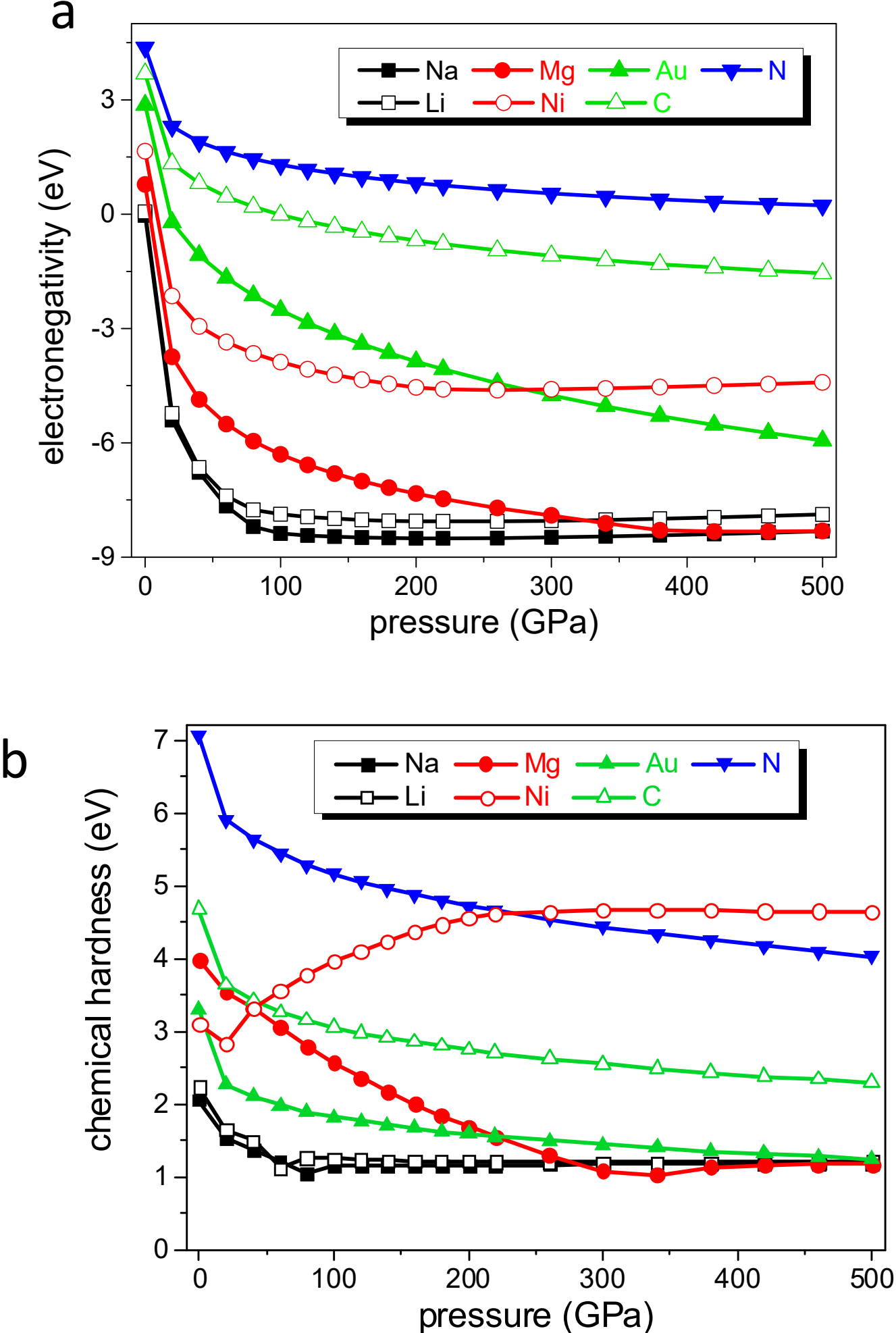

**Fig. 1. Electronegativity (a) and chemical hardness (b) as a function of pressure.**

The lowest chemical hardnesses at zero pressure belong to alkali metals, followed bt Ce and other f-elements, in the order of increasing values: Cs,Fr→Rb→Na→Li→K→Ce→... At 500 GPa, group-11 metals, actinium and magnesium join the list, again in order of increasing values:

Cu→Ag→Ac→Mg→ Rb, Na→Cs→Li→La, In, Al→...

The pressure derivative of chemical hardness equals half the difference between ionization and affinity volumes, $(V_i-V_a)/2$ and for most elements it is negative as shown in Fig S5. For a few extremely electropositive elements, such as Li, Na and Mg, there is a minimum of the chemical hardness at 60, 80 and 340 GPa, respectively. After this minimum, their chemical hardness shows slight increase with pressure: for example, Na at 500 GPa has $V_i$ = -1.78 Å$^3$ and $V_a$ = -1.89 Å$^3$. At these pressures, diffuse valence orbitals of these elements experience strong repulsion from the orbitals of the neighboring atoms, which in the He-matrix method is usually accompanied by valence electron delocalization over the He matrix and an affinity lower than that of He. In the solid forms of these elements this leads to the formation of nonmetallic or poorly metallic electride states [35,36] .

## 3. Electronic configuration and its effects on electronegativities and chemical hardnesses

Under pressure, the electronic configuration of the atoms can change because different orbitals respond differently to pressure. Occupation of orbitals with higher angular momentum will usually lower the enthalpy at sufficiently high pressures. For this reason, high pressure causes ubiquitous electronic transitions, such as s→p and s→d. Thus, it is a general rule that orbital energies and occupancies are rearranged at high pressure, which implies a rearrangement of the Periodic Table. Let us discuss some consequences of this.

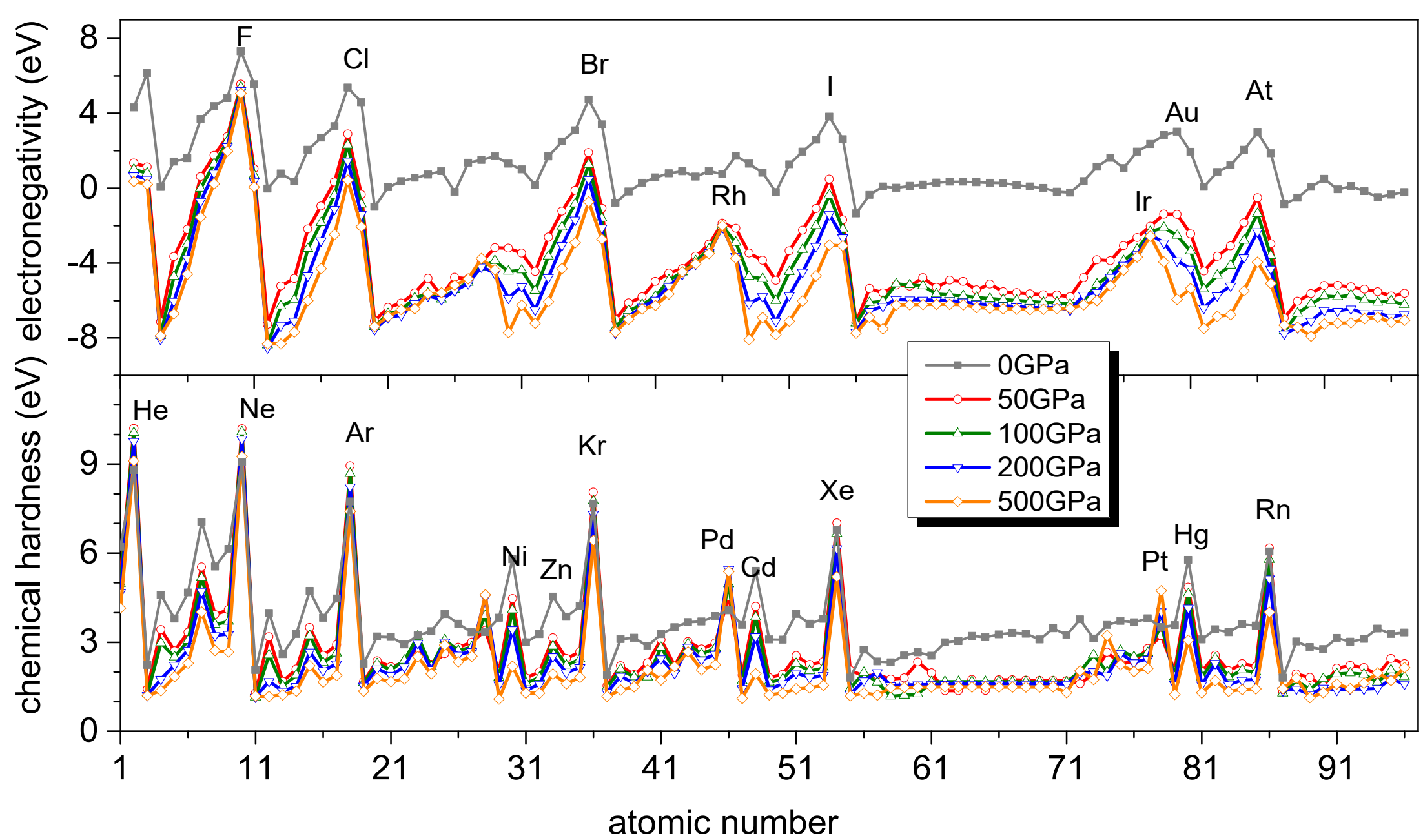

**Fig. 2. Electronegativity and chemical hardness as a function of atomic number at 0, 50, 100, 200 and 500 GPa.**

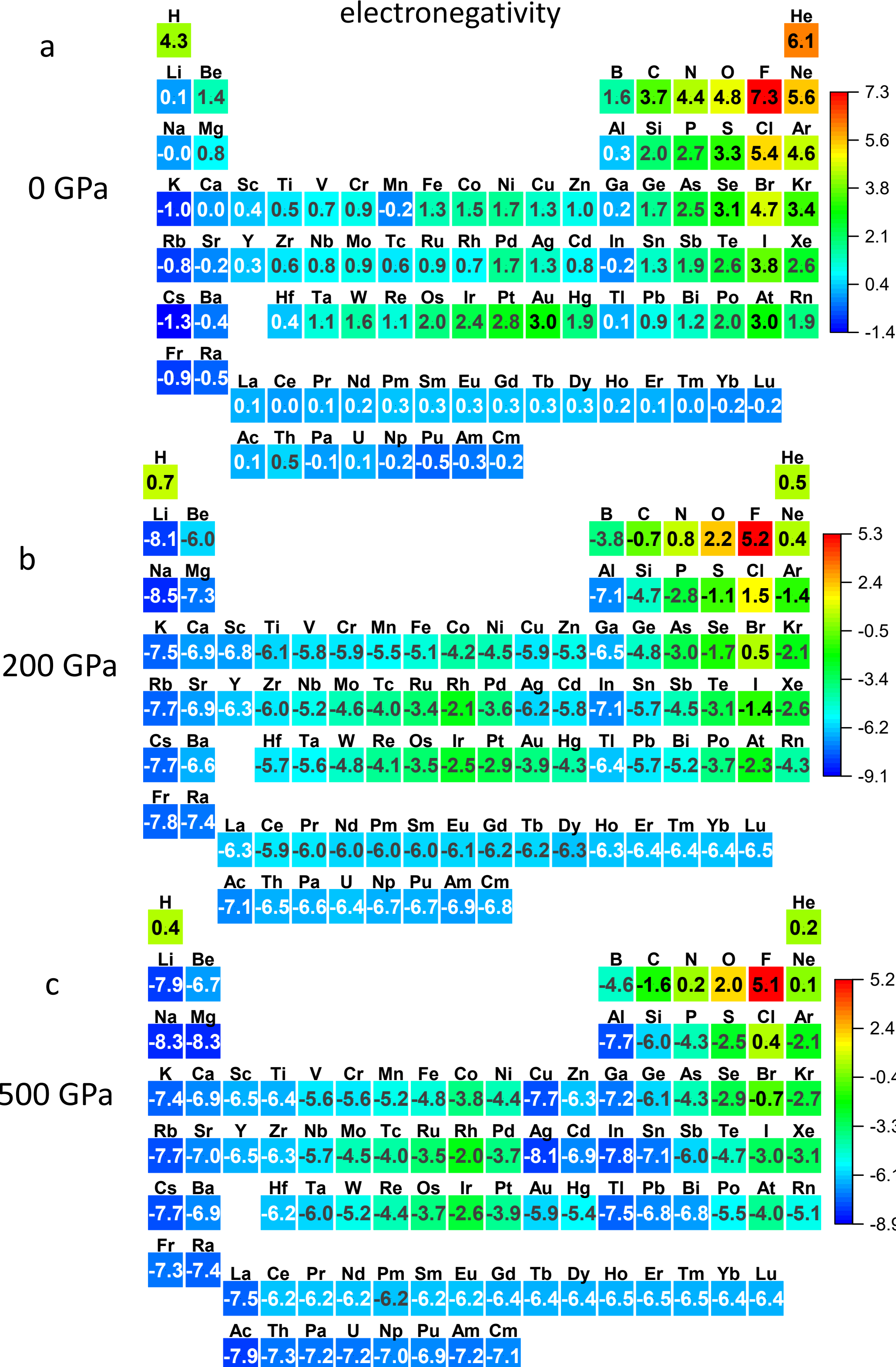

**Fig. 3. Periodic table of electronegativity at (a) 0 GPa, (b) 200 GPa and (c) 500 GPa.** Supplementary Materials contain our calculated electronegativities and chemical hardnesses of the elements, as well as their electronic configurations, magnetic moments, and non-bonded volumes at different pressures. This information provides a coherent framework to analyze the effect of pressure on atoms and chemical bonds.

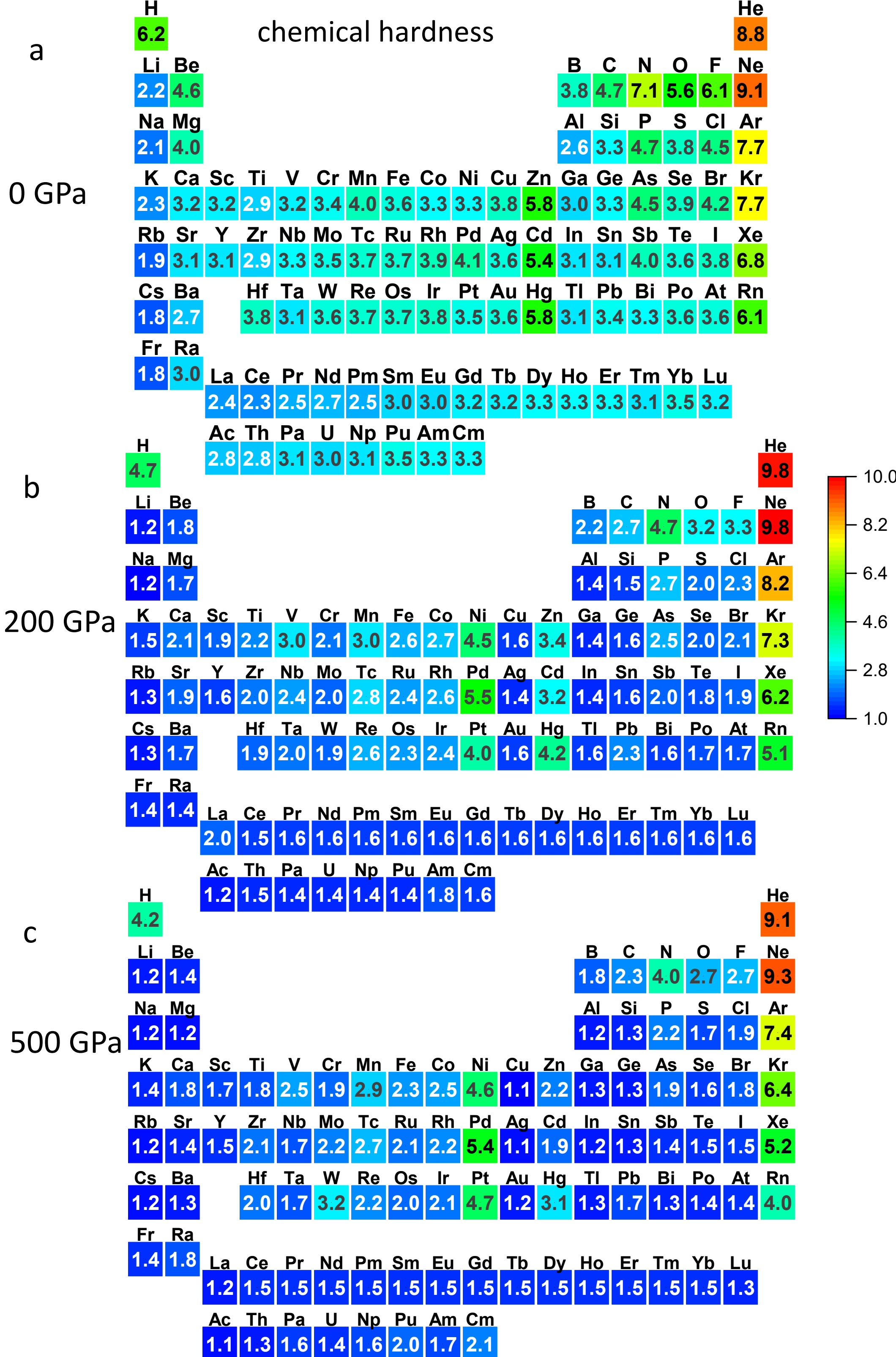

**Fig. 4. Periodic table of chemical hardness at (a) 0 GPa, (b) 200 GPa and (c) 500 GPa.**

(1) For light alkali and alkali earth metals in 2nd and 3rd periods, *s-p* interactions are essential. At low pressures, alkali earths have relatively high hardness due to $s^2$ shell and an energy gap between *s* and *p* orbitals. However, as pressure increases, the gap closes, the occupancy of *p*-orbitals increases at the expense of s-orbitals, and the chemical hardness of alkali earth metals Be and Mg becomes comparably low to that of Li and Na.

(2) Long-period elements, in particular group I (alkali metals), II (alkali earth metals) and X (nickel-group metals) display clear *s-d* orbital transfer, both in theory and experiment[22,37-41]. At zero pressure, $(n+1)s$ electrons have lower energy than $nd$ orbitals, but at high pressures this reverses and electrons will prefer to occupy the $nd$ orbitals rather than $(n+1)s$.

As shown in Fig. 2, Table S1 and S2, the Periodic Law is overall fulfilled even at high pressure, but the long-period blocks are rearranged due to the *s-d* transfer. In the 4$^{th}$ and later periods, group 1-2 elements (K, Rb, Cs, Ca, Sr, Ba) join the *d*-block under pressure, while elements of groups 11-12 (Cu, Ag, Zn, Cd, and to a lesser extent Au and Hg) under pressure behave more similar to alkali and alkali earth metals. New *d*-block elements (heavy alkali and alkali earth elements) are no longer the most electropositive. Because of the widening gap between $nd$ and $(n+1)s$ orbitals and a closed $d^{10}$-shell (formed as a result of $d^8s^2 \rightarrow d^{10}$ transition under pressure), Ni-group elements will behave more similar to noble gases, possessing low affinity and high chemical hardness at 500 GPa. For example, Pd has slightly higher hardness (5.38 eV) than Xe (5.20 eV). Fe- and Co-group elements, one or two electrons short of the Ni-configuration, will have high electronegativity: Fe will have the electronegativity of -4.84 eV comparable to Te (-4.68 eV), while Rh will have a higher electronegativity (-2.00 eV) than I (-3.03 eV). Thus Fe- and Co-group elements will acquire features of anion-forming elements at high pressure. This could have explained the unusual charge transfer from Xe to Fe in a high-pressure Xe-Fe alloy [42], but according to our results Xe remains slightly more electronegative than Fe. Similarly, Cu- and Zn-group elements, which have one or two electrons on top of the Ni-configuration, will become strong electron donors with increased reactivity: for example, Cu will have lower electronegativity (-7.71 eV) than K (-7.38 eV) at 500 GPa; also Cu has the lowest chemical hardness among all elements. To sum up, at high pressure, Ni-group atoms acquire a nearly closed-shell state, while Fe- and Co-group atoms will become stronger acceptors and Cu- and Zn-group elements–stronger donors of the electrons than at normal conditions.

(3) Under pressure Na is the most electropositive element, and the most electronegative one is still F. It is interesting to note that, contrary to common thinking that metals are the best donors of electrons and non-metals tend to be electron acceptors, high-pressure elemental Na is an insulating electride [43] and the strongest electron donor (i.e. the strongest reducing agent), unlike metallic Na at normal pressure. This is easy to understand: valence electrons in electrides, localized in the voids of the structure, are only weakly bound to the atoms (and furthermore, in high-pressure electrides may be viewed as expelled from the atoms).

Our electronegativity and chemical hardness scales can be used for explaining and predicting many phenomena, including the formation of new chemical compounds under pressure. The electronegativity and chemical hardness will have very different values and trends at high pressure, and their complex interplay creates additional complexity, and three new kinds of compounds appear:

In one class of compounds, decreased chemical hardness creates the opportunity for multicenter covalent bonding or its extreme case, the metallic bonding – the latter forming when electronegativities are low. That is what happens in the exotic sodium chlorides such as $Na_3Cl$, $Na_2Cl$, $NaCl_3$, $NaCl_7$ [11], exotic magnesium oxides such as $Mg_3O_2$ and $MgO_3$ [44] and polyhydrides, including highest-temperature superconductors, such as $LaH_{10}$ [12-15] ; $H_3S$ [16,17] ; $ThH_{10}$ [18,19] . Very low chemical hardnesses of the alkali and alkali earth elements imply that atoms of the same element can easily redistribute their electrons, creating positively and negatively sites even in a pure elemental crystal – as in host-guest structures known[45] at high pressures for such elements as Na, K, Rb, Ca, Sr, Ba, and recently predicted at ultrahigh pressures for Al [46].

The second class of compounds contains cations in unusually high valence states, stabilized by pressure. Recall that electronegativity as a function of the number N of electrons can be calculated as: $\chi(N) = \chi_0 + 2\eta_0 N$ , where $\chi_0$ and $\eta_0$ are the electronegativity and chemical hardness of the neutral atom. Let us take the example of the Cs-F system. At zero pressure, electronegativities of neutral Cs and F are -1.35 and 7.31 eV, respectively, indicating favorability of ionic bonding. At the same time, $Cs^{5+}$ has electronegativity of 16.65 eV – indeed, ionization or any other involvement of the inner electrons of Cs atom in chemical bonding is unthinkable at normal conditions. The situation changes dramatically with pressure, and at 200 GPa, $Cs^{5+}$ has the electronegativity of 4.83 eV, lower than that of neutral F (5.23 eV). Indeed, detailed theoretical calculations predict [9,10] the formation of multivalent states of Cs in such fluorides as $CsF_3$ and $CsF_5$ under pressure.

The third class of compounds, not formed at normal conditions, are stabilized by the great changes of electronegativity with pressure. Indeed, electronegativity difference adds an extra bond stabilization[47], and Miedema's model [48] of stability of intermetallides includes it as a central ingredient. Let us take the instructive example of the Mg-Fe system, which at normal conditions displays no stable compounds (and even no miscibility), due to small electronegativity difference (0.57 eV from our data). Under pressure, Mg becomes strongly electropositive, while Fe turns into an electron acceptor. At 200 GPa the electronegativity difference increases almost fourfold, to 2.25 eV, and numerous stable and strongly exothermic compounds appear (Fig. S3). MgFe is stable in the widest range of chemical potentials, and corresponds to the ideal situation of Mg donating two electrons and Fe accepting them to attain closed-shell $d^{10}$ electron configuration (adopted by Ni under pressure). Our predictions explain experimental observations[44] and theoretical prediction [49,50], which saw a great increase of miscibility in the Mg-Fe system at pressures above 100 GPa and stabilization of stoichiometric $Mg_xFe_y$ compounds among which MgFe and $Mg_5Fe_3$ have the lowest enthalpies of formation as shown in supplementary Fig. S3a. Another example is the Cu-B system: at 1 atm the electronegativity difference is very small (0.29 eV) and there are no stable bulk copper borides. With increasing pressure the electronegativity difference increases (reaching 1.00 eV at 50 GPa, 2.10 eV at 200 GPa, 3.11 eV at 500 GPa) – and stable copper boride $Cu_2B$ is predicted to appear already at 100 GPa[51]. Our electronegativities allow one to anticipate and explain the emergence of many novel compounds of this kind. Based on our tables of electronegativities, one can easily anticipate the formation of stable Na-Rh and Na-Fe compounds under pressure (at normal conditions, both systems have no stable compounds): the electronegativity difference between Na and Rh (Na and Fe) increases from 0.78 (1.38) eV at zero pressure to 6.41 (3.42) eV at 200 GPa. These expectations are indeed correct, see Fig. S3b,c.

## IV. Discussion

We have systematically explored the variation of the chemical properties of the atoms (electronegativity, chemical hardness, ionization potential, electron affinity) under pressure. In a general and systematic way, we explain many observed and predicted high-pressure chemical anomalies. Recently, Allen electronegativities (which are defined as average energies of valence electrons) were tabulated for all elements under pressure [23], but this attempt ignored the $PV$-term, which plays an essential role in high-pressure effects. The results were inexplicable: e.g., Sc became the most electropositive element [22]. Here we fully took all pressure effects (including *PV*) into account in our redefinition and detailed calculations of Mulliken electronegativities of the elements under pressure, and obtained trends and quantitative results that can be easily explained and which, in turn, explain many unusual phenomena under pressure. As comparison, we found that the lack of PV-term is unphysical and distorts both the absolute values and differences of electronegativities. For example, the difference of electronegativities of H and Br is - 1.1 eV, but without the PV term it would be 0.04 eV.

We expect other electronegativity scales to be defined and tabulated at high pressure in near future, but a correctly defined electronegativity must take into account all pressure-related effects, first and foremost the *PV*-term. For example, the new thermochemical electronegativity scale[52] is promising, and can be naturally extended to high pressure.

Pressure affects chemical behavior of the elements in several ways, related to both the decreased volume (increased electron density) and changes of the electronic structure. Let us discuss them in detail.

First, compression increases the electron density of the atoms. It is known that at very high densities kinetic energy dominates and Thomas-Fermi approximation (modeling the kinetic energy by the expression for the free electron gas of the same density) works well. We recall that Thomas-Fermi approximation predicts atoms without shell structure. This means that at sufficiently high pressures we can expect the disappearance of outer electronic shells (deeper shells disappearing at progressively higher pressures) and increasing violations and then disappearance of the Periodic Law. Heavy elements achieve a similar result at lower pressures: they have less pronounced shell structure because of decreasing energy differences between electronic shells and subshells at high principal quantum numbers. In other words, heavy elements have smaller energy differences between different valence orbitals, and as a consequence their properties show less variation than those of light elements, and their solid forms are mostly metallic. This explains the rule that under pressure elements behave like their heavier analogs from the same group at lower pressures (e.g. compressed silicon is similar to germanium and tin at normal pressure). Relativistic effects (other aspects of which are discussed below) result in further overall compression of the atoms, and this effect is strong for heavy and superheavy elements. One of its consequences is the non-inertness of the heaviest noble gas oganesson (element #118) [53,54] .

Second, electronegativities and chemical hardnesses for most elements decrease with pressure, and these trends are consistent with the universal metallization under pressure (lithium and sodium, which transform from metallic to non-metallic under pressure, are not exceptions – on further increase of pressure they eventually transform back to metallic states). Strong decrease of electronegativity and of chemical hardness renders heavy noble gases chemically more active under pressure. For a few elements (Ni-group elements – Ni, Pd, Pt) chemical hardness increases with pressure and becomes comparable to that of noble gases. This is a consequence of the closed $d^{10}$-shell acquired by these elements under pressure (due to $s^2d^8 \rightarrow d^{10}$ transition).

Third, under pressure there is a general preference for orbitals with a higher angular momentum – as a result, compressed atoms show s→p and s→d transitions. Altered electronic structure results in a rearrangement of the properties of the elements: for example, heavy alkali and alkali earth metals become d-elements under pressure, and acquire higher electronegativities than Na and Mg (which at 500 GPa are the most electropositive elements). At the same time Ni, Pd and Pt acquire a filled $d^{10}$-shell and become more inert and somewhat similar to noble gases, while and Fe- and Co-group elements possess quite high electronegativity and can serve as anions. Coinage metals Cu, Ag, and to a lesser extent (because of relativistic effects) Au behave like alkali metals, Zn-group metals become highly electropositive and behave like alkali earths (again, due to relativistic stabilization of s-electrons in heavy elements this effect in weaker for Hg). All of these changes of the Aufbau principle can be viewed as violations of the Periodic Law, and one can recall that relativistic effects in superheavy elements also produce violations of the Periodic Law. Relativistic effects, due to near-light speed of $1s^2$ electrons in heavy atoms, lead to strong compression and stabilization of all s-orbitals; d- and f-orbitals, now screened by s-orbitals, are pushed away from the atomic core. As a result of relativistic stabilization of the outer s-orbitals, Cn (element #112), the superheavy analogue of Hg, was shown to be chemically quite inert [55]. Both high pressure and relativistic effects lead to increased electron density and changes in the electronic structure, but the effects are not identical: while both effects increase the electron density of the atom, pressure destabilizes s-orbitals and stabilizes d- and f-orbitals, while relativistic effects do exactly the opposite.

## METHODS

Structure relaxations and enthalpy calculations were performed using density functional theory (DFT) within the Perdew-Burke-Ernzerhof (PBE) functional[56] in the framework of the all-electron projector augmented wave (PAW) method[57] as implemented in the VASP code[58]. We used a plane-wave kinetic energy cutoff of 1000 eV, and the Brillouin zone was sampled with a resolution of $2\pi \times 0.03$ Å$^{-1}$, which showed excellent convergence of the energy differences, stress tensors and structural parameters. All calculations included scalar relativistic effects. Search for stable Mg-Fe compounds was performed with the USPEX code [59-61] using the PBE functional and VASP code.

## ACKNOWLEDGMENTS

This work was supported by NSFC (Grants No. 21803033, No. 12174200, No. 52025026, No.11874224, No. 52090020), Young Elite Scientists Sponsorship Program by Tianjin (Grant No. TJSQNTJ-2018-18), Nature Science Foundation of Tianjin (Grant No. 20JCYBJC01530) and the foundation support from Laboratory of Computational Physics (No. 6142A05200401) and United Laboratory of High-Pressure Physics Earthquake Science. The calculations were performed at Tianhe II in Guangzhou. A.R.O. acknowledges funding from Russian Science Foundation (grant 19-72-30043) and insightful discussions with Y. T. Oganessian, R. Hoffmann and M. Rahm.